\documentclass[aps,prb,twocolumn,superscriptaddress]{revtex4}

\usepackage{graphicx}

\begin{document}


\title{Non-local Control of the Kondo Effect in a Double Quantum Dot - Quantum Wire
Coupled System}

\author{S. Sasaki}
\email[]{satoshi@nttbrl.jp}
\affiliation{NTT Basic Research Laboratories, NTT Corporation, Atsugi, Kanagawa 243-0198, Japan}

\author{S. Kang}
\affiliation{Institute of Physics, University of Tsukuba, Tsukuba, Ibaraki 305-8571, Japan}
\affiliation{CREST-JST, Honmachi, Kawaguchi, Saitama 331-0012, Japan}

\author{K. Kitagawa}
\affiliation{Faculty of Science, Tokyo University of Science, Kagurazaka, 
Tokyo 162-8601, Japan}
\affiliation{CREST-JST, Honmachi, Kawaguchi, Saitama 331-0012, Japan}

\author{M. Yamaguchi}
\affiliation{NTT Basic Research Laboratories, NTT Corporation, Atsugi, Kanagawa 243-0198, Japan}
\affiliation{CREST-JST, Honmachi, Kawaguchi, Saitama 331-0012, Japan}

\author{S.~Miyashita}
\affiliation{NTT Advanced Technology Corporation, Atsugi, Kanagawa 243-0198, Japan}

\author{T.~Maruyama}
\affiliation{NTT Advanced Technology Corporation, Atsugi, Kanagawa 243-0198, Japan}

\author{H.~Tamura}
\affiliation{NTT Basic Research Laboratories, NTT Corporation, Atsugi, Kanagawa 243-0198, Japan}
\affiliation{CREST-JST, Honmachi, Kawaguchi, Saitama 331-0012, Japan}

\author{T. Akazaki}
\affiliation{NTT Basic Research Laboratories, NTT Corporation, Atsugi, Kanagawa 243-0198, Japan}
\affiliation{CREST-JST, Honmachi, Kawaguchi, Saitama 331-0012, Japan}

\author{Y. Hirayama}
\affiliation{NTT Basic Research Laboratories, NTT Corporation, Atsugi, Kanagawa 243-0198, Japan}
\affiliation{SORST-JST, Honmachi, Kawaguchi, Saitama 331-0012, Japan}

\author{H. Takayanagi}
\affiliation{NTT Basic Research Laboratories, NTT Corporation, Atsugi, Kanagawa 243-0198, Japan}
\affiliation{Faculty of Science, Tokyo University of Science, Kagurazaka, 
Tokyo 162-8601, Japan}
\affiliation{CREST-JST, Honmachi, Kawaguchi, Saitama 331-0012, Japan}

\date{\today}

\begin{abstract}
We have performed low-temperature transport measurements on 
a double quantum dot-quantum wire coupled device and demonstrated
non-local control of the Kondo effect in one dot by 
manipulating the electronic spin states of the other.
We discuss the modulation of the local density of states in the wire region
due to the Fano-Kondo antiresonance,
and the Ruderman-Kittel-Kasuya-Yoshida (RKKY) exchange interaction
as the mechanisms responsible for the observed features.
\end{abstract} 


\maketitle

\newcommand{\tk}{T_{\rm K}}
\newcommand{\vsd}{V_{\rm sd}}
\newcommand{\nr}{N_{\rm R}}
\newcommand{\nl}{N_{\rm L}}
\newcommand{\vpl}{V_{\rm pL}}
\newcommand{\vpr}{V_{\rm pR}}
\newcommand{\vsl}{V_{\rm sL}}
\newcommand{\vdl}{V_{\rm dL}}

\begin{figure}
\includegraphics{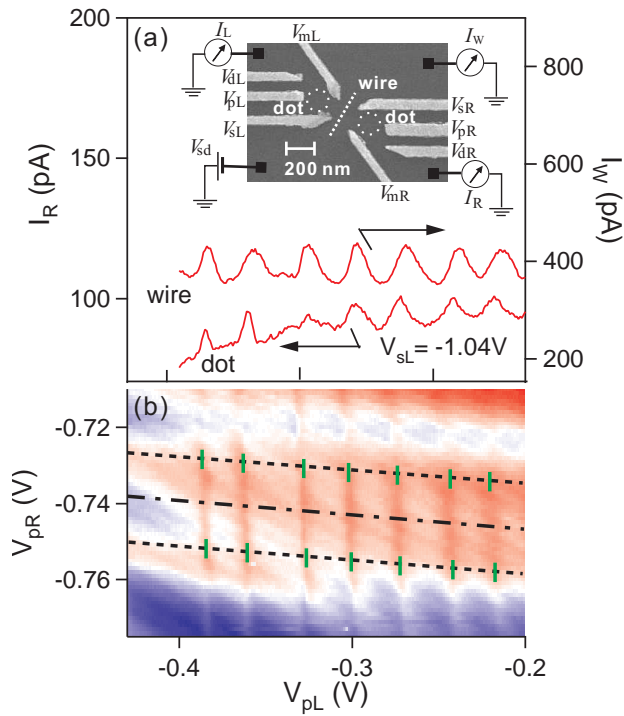}%
\caption{(a) The current through the right QD, $I_{\rm R}$, and that
through the QW, $I_{\rm W}$, in response to excitation source-drain voltage
$V_{\rm ac} = 3 \mu$V as a function of $\vpl$. $\vpr$ is changed at the same time
to trace the dash-dotted line shown in Fig.~\ref{f1}(b).
$I_{\rm R}$ and $I_{\rm W}$ are measured separately,
but with the same gate voltage conditions, which determine the wire conductance to be
around $4e^2/h$. The inset shows a
scanning electron micrograph of the double QD-QW coupled device together
with a schematic of the measurement setup. Dotted lines indicate the positions
of the QDs and QW. 
(b) Color-scale plot of the conductance through the right QD as a function
of $V_{\rm pL}$ and $V_{\rm pR}$ in the weak-coupling regime
with $V_{\rm sL} = -1.04$~V. Blue corresponds to 10~$\mu$S, white to 26~$\mu$S,
and red to 42~$\mu$S. Dotted lines denote Coulomb oscillation peaks
for the right QD, and green vertical bars denote resonances
of the left QD states with the chemical potential of the lead. 
Faint diagonal lines originate from the measurement setup.
\label{f1}}
\end{figure}

Semiconductor quantum dots (QDs) have attracted growing interest over the past years 
because of their controllable electronic states, 
and are often referred to as artificial atoms. 
One aspect of the QD as a single magnetic impurity culminated in the realization 
of an artificial Kondo system \cite{Glaz88,Ng,GGNat,Sara,Schm98,wil00}. 
When the QD has an odd number, $N$, of electrons, it normally has spin $S=1/2$. 
Then, the Kondo effect occurs when the temperature is below the Kondo temperature, $\tk$,
and manifests itself as an increase in the conductance at Coulomb blockade regions
and as a zero-bias peak in the differential conductance vs. source-drain bias characteristics. These features are observed for the conventional geometry where the source and drain electrodes 
are attached to both ends of the QD, {\it i.e.}, the only current path is 
through the ``magnetic impurity'' (embedded geometry). 
On the other hand, one can realize side-coupled geometry where a QD is attached to 
the side of another quantum wire (QW) without direct coupling to a source 
or drain electrode.
In such a case, the Fano resonance appears in the conductance of the QW
due to the interference between 
a localized state in the QD and the continuum in the QW \cite{Johnson,Koba04}. 
When the QD has odd $N$, the Kondo ``cloud'' suppresses the conductance of the QW
in the region between two Fano resonances, 
which is referred to as the Fano-Kondo antiresonance 
\cite{Kang01,Torio02,Maru04,Sato05}. 
An interesting spin correlation can be investigated when two QDs, 
each carrying spin 1/2, are coupled to the QW. 
In this case, indirect exchange interaction is expected between the two localized spins 
mediated by conduction electrons in the QW, which is known as the
Ruderman-Kittel-Kasuya-Yoshida (RKKY) interaction \cite{Tamura04,Tamura05,Craig,Simon05,SimonPRL05,Vavi05}.
The RKKY interaction may also be of great importance
in the context of quantum information
processing as a means to correlate remote spin quantum bits
connected to a common conduction channel.

In this Rapid Communication, we present transport measurement 
on two QDs coupled to the opposing sides of a quasi-one-dimensional QW.
This system offers smaller inter-dot distance in nominal device geometry
than the previous experiment by Craig {\it et al.} \cite{Craig},
and is advantageous for studying the RKKY interaction, which decays on the
length scale of the Fermi wavelength.
Moreover, the ``effective'' inter-dot distance can be tuned to zero
in an ideal case of well defined QW and zero relative distance
along the wire direction \cite{Tamura04,Tamura05}.
We demonstrate non-local control of the Kondo effect in one dot by 
tuning the spin states of the other, which suggests the RKKY interaction
as an underlying mechanism.

The inset to Fig.~\ref{f1}(a) shows the double QD - QW coupled device fabricated 
by depositing Schottky gates on a GaAs/Al$_x$Ga$_{1-x}$As heterostructure 
having two-dimensional electron gas 90 nm below the surface. 
Two QDs and a QW between them are formed (shown by dotted lines)
by biasing the eight gates. 
To avoid formation of another QD in the central QW region, 
$V_{\rm sR}$ and $V_{\rm sL}$ are biased more negatively than 
$V_{\rm mR}$ and $V_{\rm mL}$.
Transport measurement is performed with a standard lock-in technique 
at temperature $T \sim$100~mK in a dilution refrigerator. 
The source-drain bias $V_{\rm sd} = V_{\rm dc} + V_{\rm ac}$, 
where $V_{\rm dc}$ is a DC offset, and AC excitation (about 13 Hz) 
$V_{\rm ac} = 3$~$\mu$V. 

Figure \ref{f1}(b) shows a color-scale plot of the linear conductance
through the right QD as a function of $V_{\rm pL}$ and $V_{\rm pR}$.
Attention is paid to completely pinch off the exit tunnel barrier of the left QD
by applying sufficiently negative voltage, $V_{\rm dL}$.
Therefore, the only current path is through the right QD.
This is done to avoid a spurious
mirror effect, which occurs in the presence of multiple current paths \cite{Richter}. 
The gate voltage that controls the coupling of the left QD with the
rest of the system, $V_{\rm sL} = -1.04$~V, is adjusted in the weak-coupling regime.
The two dotted lines designate Coulomb oscillation peaks for the right QD,
and the region between them is a Kondo valley where the number of
electrons in the right QD, $\nr$, is odd. 
The conductance in the Kondo valley is enhanced to almost the same value as
that at the Coulomb oscillation peaks, indicating that the measurement temperature is
well below $\tk$.
Conductance maxima appear, forming almost vertical ridges as $\vpl$ is swept.
When we monitor the current through the QW only, $I_{\rm W}$, 
instead of that through the right QD, $I_{\rm R}$, as a function of $\vpl$,
the Fano resonance associated with the left QD is observed as shown in Fig.~\ref{f1}(a).
Here, the Fano's asymmetric parameter, $q$, seems to be large, 
producing almost symmetric peaks at the resonances for the present conditions \cite{com}.
$I_{\rm R}$ and $I_{\rm W}$ exhibit almost in-phase oscillations
with the period of 20 $\sim$ 30~mV in $\vpl$.
This period is also consistent with the separation between Coulomb oscillation
peaks observed in the direct current through the left QD, $I_{\rm L}$, 
which flows with more positive $\vdl$. 
Therefore, we assign the ridges shown in Fig.~\ref{f1}(b)
to resonances of the left QD states with the chemical potential of the lead.
The Fano resonance modulates the conductance of the QW,
and $I_{\rm R}$, which flows via the QW, is also modulated.
The ridges are not observed when $\vsl$ is smaller than about $-1.2$~V 
because the left QD is decoupled from the rest of the system.
Although the Kondo valley conductance is somewhat reduced from the original value
at $\vpl = 0$, the Kondo effect still remains between the ridges.
Please note that the two leftmost ridges are separated from the others
by a larger gap, suggesting a spin pair where up-spin and down-spin electrons
consecutively occupy the same orbital state.
However, there is no noticeable difference in the conductance between 
even and odd $\nl$ regions, where $\nl$ is the number of electrons in the left QD.

\begin{figure}
\includegraphics{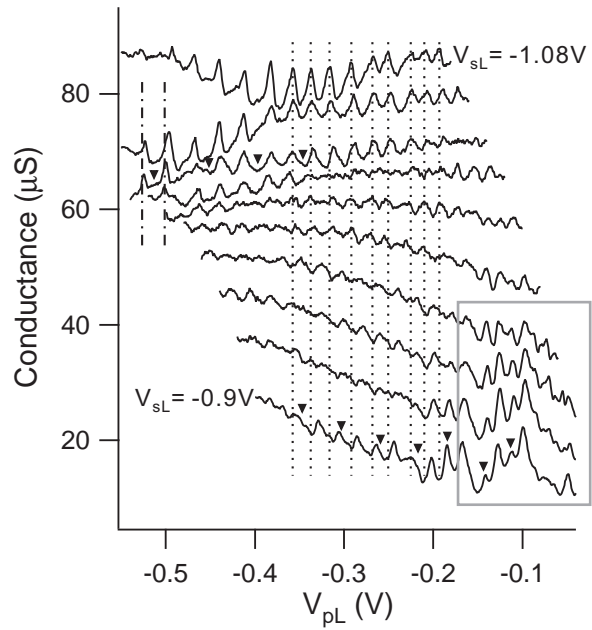}%
\caption{Conductance profiles taken along the mid Kondo valley of the right QD 
with the coupling gate voltage $\vsl$ changed between -0.9 and -1.08~V in 20~mV steps.  
Curves are offset horizontally to compensate for the electrostatic energy
shift due to $\vsl$ and also vertically for clarity.
The scales on the axes apply to the lowermost $\vsl = -0.9$~V trace.
The dotted lines are guides for the eye, and the two dash-dotted lines on the left
denote spin-pair peaks.
Odd $\nl$ regions are marked with solid triangles.
The boxed region is plotted in more detail in Fig.~\ref{f3}.
\label{f4}}
\end{figure}

Next, we increase the coupling of the left QD by increasing $\vsl$.
Figure~\ref{f4} shows evolution of the conductance profiles taken along the
mid Kondo valley of the right QD (similar to Fig.~\ref{f1}(a)) 
when $\vsl$ is changed from $-1.08$~V
(weak-coupling regime) to $-0.9$~V (strong-coupling regime).
Asymmetric peaks reflecting the Fano resonances are observed in the weak-coupling regime.
We estimated the shift in $\vpl$ of the resonance peaks in the weak-coupling regime
in response to a slight shift in $\vsl$ that does not affect the Fano line shape, 
and found that the electrostatic energy shift caused by $\vsl$ is almost identical 
to that caused by $\vpl$ in our device. 
Based on this finding, we shift the traces in Fig.~\ref{f4}
horizontally so that the resonances arising from the same $\nl$ states
are aligned vertically.
Using the dotted lines as a guide, one can see that the 
resonances appearing as conductance maxima in the weak-coupling regime
change into minima in the strong-coupling regime,
just as in the previous report by Sato {\it et al} \cite{Sato05}.
We assign odd $\nl$ regions and mark them with solid triangles,
using the spin pair behavior in Fig.~\ref{f1}(b) as a reference 
(the two dashed lines in Fig.~\ref{f4}).

\begin{figure}
\includegraphics{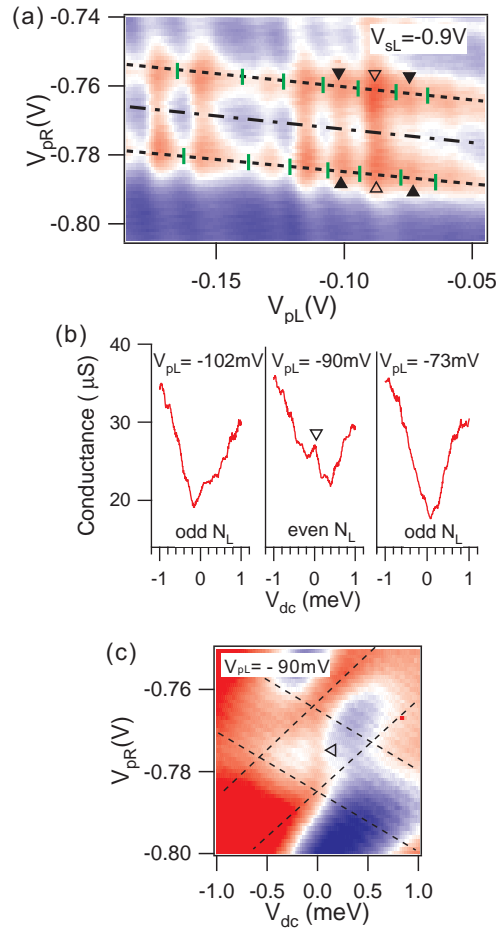}%
\caption{(a) Color-scale plot of the conductance through the right QD as a function
of $V_{\rm pL}$ and $V_{\rm pR}$ with $V_{\rm sL} = -0.9$~V 
in the strong-coupling regime. Blue corresponds to 0~$\mu$S, white to 18~$\mu$S,
and red to 36~$\mu$S. Dotted lines and green vertical bars have the same
meanings as in Fig.~\ref{f1}(b).
The open and solid triangles denote even and odd $\nl$ regions, respectively.
(b) d$I_{\rm R}$/d$V_{\rm dc}$ vs $V_{\rm dc}$ for odd $N_{\rm L}$ 
($V_{\rm pL} = -102$~mV and $-73$~mV, the solid triangles in (a)) 
and for even $N_{\rm L}$ ($V_{\rm pL}$ = $-90$~mV, the open triangle in (a)). 
$V_{\rm pR}$ is fixed at $-775$~mV in the middle of the
odd $N_{\rm R}$ Coulomb blockade valley.
The triangle for $V_{\rm pL}$ = $-90$~mV case denotes a Kondo resonance peak.
(c) Conductance of the right QD as a function of $V_{\rm dc}$ and $V_{\rm pR}$
for even $\nl$.
Blue corresponds to 10~$\mu$S, white to 25~$\mu$S,
and red to 40~$\mu$S.
The triangle denotes a Kondo ridge observed in the odd $N_{\rm R}$ Coulomb 
blockade region. The edges of the Coulomb diamonds are shown by dashed lines.
\label{f2}}
\end{figure}

Figure~\ref{f2}(a) shows a plot similar to Fig.~\ref{f1}(b), 
but with $\vsl = -0.9$~V, in the strong-coupling regime.
Here, in contrast to Fig.~\ref{f1}(b),
the Fano resonance involving the left QD with $q \simeq 0$ produces 
conductance minima (green vertical bars in Fig.~\ref{f2}(a)).
The intensity of the several conductance maxima corresponding to fixed $\nl$ 
seems to alternate between $\vpl$ = $-130$~mV and $-70$~mV. 
As shown in Fig.~\ref{f2}(b),
the zero-bias Kondo peak still remains in the d$I_{\rm R}$/d$V_{\rm dc}$ vs $V_{\rm dc}$
characteristics when $\nl$ is even ($\vpl = -90$~mV). 
The Kondo effect for this condition is also confirmed
in the Coulomb diamond characteristics shown in Fig.~\ref{f2}(c).
On the other hand, there is no Kondo peak when $\nl$ is odd 
($\vpl = -102$~mV and $-73$~mV, solid triangles in Fig.~\ref{f2}(a)).
Thus, a non-local control of the Kondo effect in the right QD is realized
by changing the number of electrons in the left QD.

\begin{figure}
\includegraphics{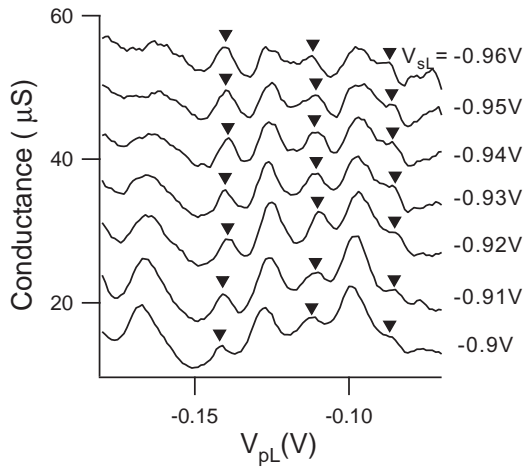}%
\caption{Conductance profiles taken along the mid Kondo valley of the right QD 
when the coupling of the left QD is changed with $\vsl$.  
Each curve is offset by 5~$\mu$S vertically 
and by $-10$~mV horizontally with respect to the previous curve with 10~mV more $\vsl$.
The scales on the axes apply to the lowermost trace.
The solid triangles denote odd $\nl$ conductance peaks.
There is an overall shift of about 10~mV in $\vpl$ compared to Fig.~\ref{f2}
due to the different measurement time.
\label{f3}}
\end{figure}

Figure~\ref{f3} shows the $\vsl$ dependence of the 
mid Kondo valley conductance profiles 
taken from the boxed region in Fig.~\ref{f4} with smaller 10~mV steps in $\vsl$.
Dips correspond to the Fano resonances associated with the left QD, and peaks
correspond to the fixed $\nl$ regions.
The intensity of the odd $\nl$ peaks (marked with triangles) is not 
very different from that of the even $\nl$ peaks when the coupling is relatively weak
($\vsl \simeq -0.96$~V).
As $\vsl$ is made more positive ($\simeq -0.92$~V), dips due to the Fano
resonance develop and make the peaks clearer, 
while the overall conductance slightly decreases.
When $\vsl$ is further increased and the left QD is strongly coupled, 
odd $\nl$ peaks are suppressed more than even $\nl$ ones, leading to the suppression
of the Kondo effect as seen in Fig.~\ref{f2}(b).

Let us now consider the mechanisms responsible for the observed even-odd effect.
As already mentioned, the dot current, $I_{\rm R}$, is measured via the QW region, 
{\it i.e.}, the QW acts as one lead for the right QD.
When the Fano resonance involving the left QD 
modulates the conductance of the QW,
it also indirectly modulates $I_{\rm R}$.
As coupling of the left QD with the QW increases, the Fano-Kondo antiresonance effect
suppresses the conductance at the odd $\nl$ regions \cite{Sato05},
consistent with the behavior observed in Fig.~\ref{f3}.
It must be noted that the observed even-odd effect is different from a simple
series resistance modulation. If that were the case, the zero-bias Kondo
peak would still appear, but with a reduced height, in the case of odd $\nl$ too.
In reality, however, the Kondo temperature itself is changed
between even and odd $\nl$. 
It is expected that the Fano resonance (or Fano-Kondo antiresonance) 
modulates the local density of states of the QW,
as the enhanced zero-bias conductance shown in Fig.~3 in Ref.~\cite{Koba02} suggests. 
Because $\tk \propto \exp(-1/\rho J)$, where $\rho$ is the density of states
and $J$ is the coupling constant \cite{cox},
$\tk$ decreases when $\rho$ decreases.
This may be the case for the odd $\nl$ peaks in Fig.~\ref{f3}
because $\rho$ is reduced by the Fano-Kondo effect compared to even $\nl$ ones.
Usually, the Kondo effect is experimentally controlled with 
a variety of parameters such as temperature, the tunnel coupling with the leads, 
the position of the localized level relative to the Fermi energy,
source-drain bias, and magnetic field. 
The control of $\tk$ via the Fano(-Kondo)-modulated density of states represents
a totally new way to experimentally switch on/off the Kondo effect.

Another mechanism that could be responsible for the observed even-odd effect
is the RKKY interaction.
When both $\nr$ and $\nl$ are odd, the RKKY interaction couples two spins 
in the two QDs via interceding conduction electrons 
either ferromagnetically or anti-ferromagnetically depending on 
the distance between them. The Kondo effect for the single QD with spin 1/2 is
suppressed both in the former case (the total spin, $S_{\rm tot} = 1$), 
corresponding to the underscreened Kondo effect, and in the latter case ($S_{\rm tot} = 0$).
The effective inter-dot distance in our device structure is considered to be
close to zero since they couple to the opposing sides of the quasi-one-dimensional QW
(having two to three conduction channels for the gate voltage conditions explored) 
with a small relative distance along the wire direction \cite{Tamura04, Tamura05}.
Hence, the ferromagnetic coupling may be realized, and relatively large interaction
strength is expected.
However, this RKKY mechanism is inseparable from the Fano-Kondo-modulated 
density of states scenario described above, and presumably both mechanisms
contribute to the observed suppression of the Kondo effect when the remote
QD also carries spin 1/2. 

In summary, we have presented transport measurement on a double QD-QW 
coupled system and demonstrated non-local control of 
the Kondo effect in one QD by manipulating the spin states of the other.
As for the responsible mechanisms, we consider modulation of the Kondo 
temperature by the modulated local density of states of the QW due to 
the Fano-Kondo antiresonance, 
and the RKKY interaction between the two QDs.

\begin{acknowledgments}
This work is financially supported by Grant-in-Aid for
Scientific Research from the Japan Society for the Promotion of Science.
\end{acknowledgments}


\begin{thebibliography}{99}
\bibitem{com}
Transmission line shape is given as $G (\epsilon) \propto (\epsilon + q)^2 / 
(\epsilon^2+1)$, where $\epsilon$ is the detuning from the resonance position 
normalized with the level broadening, and $q$ is Fano's asymmetric parameter.
\end{thebibliography}


\end{document}